\documentclass[]{article}
\def\*#1{\boldsymbol{#1}}

\usepackage{PRIMEarxiv}

\usepackage[utf8]{inputenc} 
\usepackage[T1]{fontenc}    
\usepackage{url}            
\usepackage{booktabs}       
\usepackage{amsfonts}       
\usepackage{nicefrac}       
\usepackage{microtype}      
\usepackage{lipsum}
\usepackage{fancyhdr}       
\usepackage{graphicx}       
\graphicspath{{media/}}     

\usepackage{amsmath}
\usepackage{amssymb}
\usepackage{amsthm}
\usepackage{bbm}
\usepackage[square,authoryear]{natbib}
\usepackage[dvipsnames]{xcolor}
\usepackage[colorlinks=true,
	linkcolor=violet,
	citecolor=blue]{hyperref}

\usepackage{comment}

\usepackage{algorithm}
\usepackage{algpseudocode}
\usepackage{algorithmicx}

\usepackage{caption}
\usepackage{subcaption}

\usepackage{listings}
\usepackage{xcolor}

\definecolor{codegreen}{rgb}{0,0.6,0}
\definecolor{codegray}{rgb}{0.5,0.5,0.5}
\definecolor{codepurple}{rgb}{0.58,0,0.82}
\definecolor{backcolour}{rgb}{0.95,0.95,0.92}

\lstdefinestyle{mystyle}{
    backgroundcolor=\color{backcolour},   
    commentstyle=\color{codegreen},
    keywordstyle=\color{magenta},
    numberstyle=\tiny\color{codegray},
    stringstyle=\color{codepurple},
    basicstyle=\ttfamily\footnotesize,
    breakatwhitespace=false,         
    breaklines=true,                 
    captionpos=b,                    
    keepspaces=true,                 
    numbers=left,                    
    numbersep=5pt,                  
    showspaces=false,                
    showstringspaces=false,
    showtabs=false,                  
    tabsize=2
}

\title{Spatial Transcriptomics Dimensionality Reduction \\ using Wavelet Bases} 

\author{
  Zhuoyan Xu \\
  Department of Statistics \\
  University of Wisconsin - Madison \\
  \texttt{zhuoyan.xu@wisc.edu} \\
   \And
  Kris Sankaran \\
  Department of Statistics \\
  University of Wisconsin - Madison \\
  \texttt{ksankaran@wisc.edu} \\
}

\begin{document}
\maketitle

\begin{abstract}
Spatially resolved transcriptomics (ST) measures gene expression along with the spatial coordinates of the measurements. The analysis of ST data involves significant computation complexity. In this work, we propose gene expression dimensionality reduction algorithm that retains spatial structure. We combine the wavelet transformation with matrix factorization to select spatially-varying genes. We extract a low-dimensional representation of these genes. We consider Empirical Bayes setting, imposing regularization through the prior distribution of factor genes. Additionally, We provide visualization of extracted representation genes capturing the global spatial pattern. We illustrate the performance of our methods by spatial structure recovery and gene expression reconstruction in simulation. In real data experiments, our method identifies spatial structure of gene factors and outperforms regular decomposition regarding reconstruction error. We found the connection between the fluctuation of gene patterns and wavelet technique, providing smoother visualization. We develop the package and share the workflow generating reproducible quantitative results and gene visualization. The package is available at \url{https://github.com/OliverXUZY/waveST}.
\end{abstract}


\section{Introduction}
Spatial resolved transcriptomics (ST) is a technology measuring spatial variation in gene expression. Different technological platforms support different genome coverage and spatial resolution (tissue-level measurement to subcellular measurement). Examples of ST platforms include Spatial Transcriptomics \citep{staahl2016visualization, xia2019spatial}, 10x Genomics Visium, Slide-seq \citep{rodriques2019slide}, sci-Space \citep{srivatsan2021embryo}, and MERFISH \citep{chen2015spatially}. Spatial transcriptomics allows visualization and quantitative analysis of the transcriptome with spatial resolution in individual tissue sections. A series of studies combining gene expression and spatial information has been brought to generate new insight in biological analysis \citep{kuppe2020spatial, shah2016situ, berglund2018spatial}. 

Quantification of gene expression has wide applications in transcriptomics. Understanding the spatial distribution of gene expression has helped to answer fundamental questions in developmental biology \citep{asp2019spatiotemporal, rodelsperger2021spatial}, cancer \citep{thrane2018spatially, moncada2020integrating}, and neuroscience \citep{moffitt2016high, close2021spatially}. Two widely-used methods for gene expression quantification are fluorescent in situ hybridization (FISH) and next-generation sequencing.

While these approaches have been made to measure gene expression while preserving spatial information, there are statistical challenges in analyses combining the gene and spatial information. Specifically, gene dimensionality reduction guided by spatial information is still an active area of study. In this work, we conduct a dimensionality reduction on spatially varying gene data. After transformation, gene expression over locations is a square matrix, which allows us to view it as an image. Therefore,  tools from image processing can be adapted -- our method incorporates spatial information by wavelet transformation, a multi-scale analysis decompose sequence into orthonormal series. We apply wavelet transformation for each gene expression over locations. This is a common technique in denoising the image. We thus apply techniques analogous to image analysis. We evaluate performance using reconstruction error and comparing resulting visualizations.  Our analysis pipeline build from off-the-shelf tools, and code is available for reproducing our results. We also attach visualizations in our pipeline, giving example interpretations of intermediate result. 

A natural idea in spatial transcriptomic analysis is to identify gene types across spatial locations \citep{abu2018clust}. For example, in clustering the representations in each cluster achieve dimensionality reduction under the assumption that genes within one cluster have the same type. A parallel technique is clustering based on cell type instead. General clustering methods can be combined into more sophisticated pipelines tailored toward spatial single-cell analysis. SC3 \citep{SC3package} is an ensemble clustering method. It calculates distance matrices across cell locations using the Euclidean distance, then applies spectral clustering, and assigns membership \citep{kiselev2018scmap}. To identify the cell type in each cluster, one can perform differential expression analysis between all pairs of clusters. scGeneFit uses a label-aware compression method to find marker genes \citep{dumitrascu2021optimal}. Given the cell-by-gene expression matrix and cell clustering membership, scGeneFit maps cells to lower-dimensional space where cells within the same cluster are closer. For gene dimensionality reduction, \citep{zhu2020integrative} constructs a neighborhood graph from the spatial coordinates, then applies a graph-based feature selection procedure to determine spatially varying genes. They also provide the option to infer a latent graph embedding for cells based on selected genes, applying spline models to fit the gene's expression on the latent embedding. Then they leverage the fitted coefficients to reduce the dimensionality of each gene. \citep{svensson2018spatialde} proposed pipelines using mixed-effect models incorporating spatial information. The model contains two random effect terms: a spatial variance term that parametrizes gene expression covariance by pairwise distance between samples and a noise term that models nonspatial variability. The model leverages efficient inference methods previously developed for linear mixed models, and it is computationally efficient. 

Our setting is closely aligned with the following recent works: \citep{shang2022spatially} developed SpatialPCA,  applying probabilistic PCA on ST data for dimensionality reduction. They assume data are given as a location-by-genes matrix and construct a regression model similar to factor analysis, where the prior covariance matrix of factor genes is a distance matrix constructed with a Gaussian kernel. \citep{velten2022identifying} proposed MEFISTO, combining factor analysis with the non-parametric framework of Gaussian processes to model spatio-temporal dependencies in the latent space. \citep{townes2021nonnegative} developed nonnegative spatial factorization (NSF), combining a Gaussian process prior over spatial locations and a Poisson or negative binomial likelihood for count data, identifying generalizable spatial patterns of gene expression. All these works impose spatial structure on the prior of the factor genes. While these methods offer new dimensionality reduction techniques to cluster the genes, the complex model structure and a large number of hyperparameters introduce uncertainty and noise. Instead of imposing structural assumptions on the prior of the factor genes, we impose structure on the factor gene itself. Our contributions are the following:
\begin{itemize}
    \item We propose an approach, based on techniques from matrix decomposition and image signal processing to perform gene dimensionality reduction that retains inferred spatial structure.
    
    \item  We run simulations showing that wavelet-guided dimensionality reduction performs better estimation than the singular value decomposition (SVD) under low signal-to-noise (SNR) regime. 
    
    \item We perform real data experiments to
    identify the connection between wavelet techniques and fluctuation of the gene expression, which would be useful in 
    selecting spatially related genes based on reconstruction error.
    
    \item We provide a gene extraction pipeline capturing the global information of spatially related genes. We provide smoother visualization of gene factors via wavelet methods. We develop an R package \texttt{waveST} and share the workflow generating reproducible quantitative results and gene visualization. 
\end{itemize}

\begin{figure}[!htbp] 
\begin{center} 
\includegraphics[width=14cm]{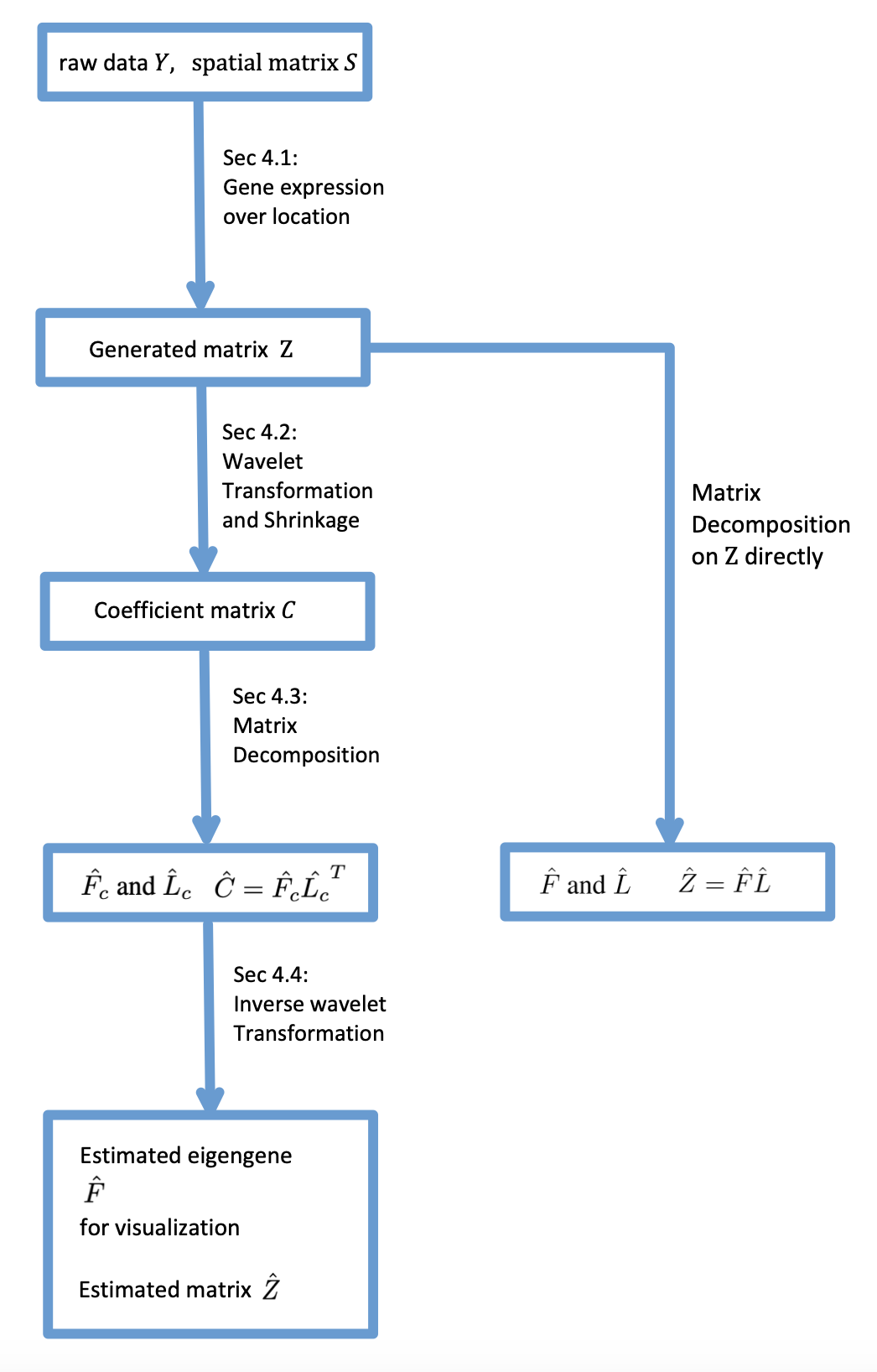}
\end{center}
\caption{A summary of the proposed workflow.} 
\label{diagram}
\end{figure}

The diagram for workflow can be seen in Figure \ref{diagram}. The paper is organized in following: In Section \ref{SecBackground} we introduce background on required techniques, including the wavelet transformation and matrix decompositions. In Section \ref{problemSetup}, we formally define our problem under this setting, and section \ref{methods} introduces our algorithms and analysis pipeline. In section \ref{Sec_sim}, we implement simulations showing the effect of wavelet transformation in reducing error. In section \ref{sectionRealData} we conduct our method on data from \citep{STdata}, showing the reconstruction error in dimension reduction and visualization of lower-dimensional representations.

\section{Background}  \label{SecBackground}
In this section, we will cover background of methods we used in data pre-processing and analysis.

\subsection{Wavelet Transformation} \label{waveBackground}
Density estimation and function approximation is a fundamental problem in statistics and machine learning. Non-parametric methods such as spline regression \citep{perperoglou2019review}, Fourier transformation \citep{cochran1967fast} and Wavelet transformation \citep{nason2008wavelet} has been used in such scenarios.

Consider a model with one predictor: $y=f(x)+\epsilon$, where $\mathbb{E}\epsilon^2 = \sigma^2$. We want to estimate $f$, the \emph{trend} of the function. We assume the predictors are ordered $x_{1}<x_{2}<\ldots<x_{n}$. We have a signal or frequency to estimate.

Consider the \emph{Haar mother} wavelet:
$$
\psi(x)=\left\{\begin{array}{cc}1 & x \in\left[0, \frac{1}{2}\right) \\~\\ -1 & x \in\left[\frac{1}{2}, 1\right) \\~\\ 0 & \text { otherwise }\end{array}\right.
$$
satisfying $\int_{-\infty}^{\infty} \psi(x) d x=0$. Unlike the Fourier basis, wavelets oscillate and decay fast, only contributing to a certain local area and zero elsewhere. We can generate wavelets from the Haar mother wavelet $\psi_{j, k}(x)=2^{j / 2} \psi\left(2^{j} x-k\right)$ for integer $j,k$. These wavelets form orthonormal sets. We can decompose the trend as $f(x)=\sum_{j=-\infty}^{\infty} \sum_{k=-\infty}^{\infty} d_{j, k} \psi_{j, k}(x)$, where $d_{j, k}=\int_{-\infty}^{\infty} f(x) \psi_{j, k}(x) d x=<f, \psi_{j, k}>$ are the wavelet coefficients.

Wavelet based methods have several advantages among other non-parametric methods, especially dealing with sparse data. One property wavelet has is localization. If a sequence has a discontinuity, this will only influence the wavelet basis around it. In contrast, for a Fourier basis consisting of sine and cosine functions at different frequencies, every basis element will interact with this discontinuity, hence influencing every Fourier coefficient.

The simplest discrete wavelet transformation calculates the difference and sums between each adjacent pair. Suppose we have one vector with length $n$, where $n$ is dyadic ($n = 2^J$). We computed $d_{J-1,k}=y_{2 k}-y_{2 k-1}$ as finest-level detail and $c_{J-1,k}=y_{2 k}+y_{2 k-1}$ as the finest-level averages. We have $\left\{d_{k}\right\}$ containing $n/2 = 2^{J-1}$ as our finest level coefficients. To obtain the next coarsest coeffients we set:
\begin{align} 
    d_{J-2, \ell}&=c_{J-1,2 \ell}-c_{J-1,2 \ell-1} \label{d} \\
    c_{J-2, \ell}&=c_{J-1,2 \ell}+c_{J-1,2 \ell-1} \label{c}
\end{align}
We continue this procedure until $j$ reaches one. The set of details $\left\{d_{k}\right\}$ across levels are our wavelet coefficients.

We use $J$ to denote the scale of wavelets, with larger $J$ we have finer scale and approximation. However, finer scale sometimes introduces more parameters to capture minor details of the sequence (overfitting). The trade-off in choosing $J$ is crucial in the wavelet transformation.

Regularization and smoothing is can be used to prevent overfitting. Regularization is usually conducted by shrinking wavelet coefficients. The concept of wavelet shrinkage was first proposed by \citep{donoho1994ideal}. The motivation behind wavelet shrinkage is straightforward. Consider empirical wavelet coefficients, the large coefficients usually contain true signal and noise, whereas the small coefficients only contain noise. The shrinkage is often used when the wavelet coefficients are assumed to be a \emph{sparse vector}. 

Wavelet shrinkage is often conducted by setting a \emph{threshold} (denoted by $\delta$) and only keeping the coefficients above the threshold. To choose a threshold, a natural metric is the squared error between estimated function and the truth: $\hat{M} = \sum_{i = 1}^n (f(x_i) - \hat{f}(x_i))^2$ and $M = \mathbb{E}\hat{M}$. \citet{donoho1994idealb} proposed a \emph{universal threshold} $\delta = \sigma\sqrt{2 \log n}$, which induced $M \le \mathcal{O}(\log n \sigma^2)$. \citet{donoho1995adapting} also proposed \emph{Sure threshold} based on Stein's (1981) unbiased risk estimator. The optimal SURE threshold can be obtained in $\mathcal{O}(n \log n)$ operations. \citet{donoho1995adapting} also noted that SURE sometimes failed when the true signal coefficients are highly sparse, they proposed a hybrid scheme called \emph{SureShrink}, combining the SURE and universal thresholds, using them depending on certain situations.

The extension of wavelet methods to 2D regularly spaced data (images) and such data in higher dimensions was proposed by \citep{mallat1989theory}. We only consider 2-D wavelet transformation since we decompose 2-D spatial gene expression data. Suppose we have $n \times n$ matrix $A$ where $n = 2^{J}$ is dyadic. A simple discrete wavelet transformation on $A$ first applies procedure (\ref{d}) and (\ref{c}) to the rows of the matrix. We then have two matrices of size $n \times \frac{n}{2}$, called $H$ and $G$. Then we apply the same procedures to both the columns of $H$ and $G$, resulting in four matrices $HH, GH, HG$, and $GG$ each of size $\frac{n}{2} \times \frac{n}{2}$. These are our finest level coefficients. $HH$ is the local average of the original matrix used for the next level procedure.

\subsection{Factor Gene} \label{subsec_factorgene}

Clustering and dimensionality reduction are widely used in genomics. In a gene-by-sample matrix, genes are often grouped into profiles, where genes from the same profile have a similar function. Statistically, we can treat them as correlated variables.
We use the term factor genes as the principal component in our gene-by-sample data. Factor genes compose the linear combinations of genes. We consider each gene as a variable, and we aim to find variables $(\boldsymbol{f_1}, \ldots, \boldsymbol{f_K})$ such that each gene $g_i$ has form $\boldsymbol{g_i} = a_{1} \boldsymbol{f_{1}}+a_{2} \boldsymbol{f_{2}}+\cdots a_{K} \boldsymbol{f_{K}}$, where $a_j$ are coefficients. 

Suppose we have gene($P$)-by-sample($N$) matrix $A \in  \mathbb{R}^{N \times P}$ with sample covariance matrix $S = \frac{1}{n}\Tilde{A}^T\Tilde{A}$, where $\Tilde{A}$ is the column centered $A$. Consider the SVD on $\Tilde{A} = U \Lambda V^T$. Then we have $SV = V\Lambda^2$. The columns of $V$ are called \emph{eigenarrays} \citep{alter2000singular}. The first $K$ columns of $Z = AV$ are the \emph{factor genes}. The factor genes capture the mutual underlying information of genes.

\subsection{Empirical Bayes Matrix Factorization} \label{EBMF}
Matrix Factorization is often used in capture factor genes. We have a formulation similar to \citep{wang2021empirical}, consider the factorization model on observed samples by gene data $Y \in \mathbb{R}^{N \times P}$ $$
Y=F L^{T} +E
$$
where $F \in \mathbb{R}^{N \times K}$ denotes the factors, $L \in \mathbb{R}^{P \times K}$ denotes the loadings for each factor, and $E \in \mathbb{R}^{N \times P}$ denotes Gaussian noise with zero mean. Typical formulations would treat factors and loadings as fixed effects and use Maximum Likelihood Estimation (MLE). In our high-dimensional setting, penalty-based regularizers are often considered. Considering a prior for $L$ and $F$ under Bayesian setting has a similar effect and several advantages over adding a regularizer alone, such as simplifying hyperparameter search and selection of the number of factors $K$. 

One feature of empirical Bayes approaches in matrix factorization proposed by \citep{bishop1999variational} is that the methods automatically select the number of factors $K$. In \citep{wang2021empirical}, the author add factors with prior one at a time, estimating priors at each step until convergence. If the computed prior of the newly added factor is almost point mass on 0, the algorithm eliminates this factor and returns.

Following \citep{wang2021empirical}, we consider \emph{Empirical Bayes} in our setting, where we set a prior with unknown parameters of $L$ and $F$. Empirical Bayes is not strictly Bayes, since the prior parameters are directly estimated by the MLE of the data. One example would be normal distribution $\mathcal{D}_F$ of $F$ and $\mathcal{D}_L$ of $L$ where coordinates are independent, which is conjugate prior. We estimate parameters in $\mathcal{D}_F$ and $\mathcal{D}_L$ by the maximize the marginal likelihood calculated by integral out $L$ and $F$. Then we computed the posterior distribution of $L$ and $F$.

\section{Problem Setup} \label{problemSetup}

Consider $Y \in \mathbb{R}^{N \times P}$, where each row represents a sample and each column represents a gene. Further, let $S \in \mathbb{R}^{N \times 2}$ store spatial coordinates of each sample. We assume there are spatially-related genes among all genes. We assume the genes fall into several profiles. The genes in the same profile have the same expression over the sampled spatial context. For each profile, we can summarize the gene expression for such group by one representative factor. Consider there are $K$ profiles, we can then form our model as:
\begin{align*}
    Y&= \sum_{k=1}^{K} \boldsymbol{f}_{k} \boldsymbol{l}_{k}^{T}+E \\
    &= F L^{T}+E
\end{align*}

where $\* f_k\in \mathbb{R}^{n}$ is the factor gene in profile $k$ capturing the gene expression pattern in that group, $\*l_k \in \mathbb{R}^{p}$ indicates the loading coefficients of each factor. We assume there is random noise $E \in \mathbb{R}^{n \times p}$ in observations and $E_{i j} \overset{\text{iid}}{\sim} N(0,1 / \tau)$.
The goal is to select spatially-related genes based on spatial information in $S$. We also extract the gene factors which capture information across all genes based on feature extraction. We aim to find a latent gene space that respects spatial structure.

\section{Methods} \label{methods}
In this section, we build up pipelines and models to achieve this goal. We adapt matrix decomposition methods to incorporate spatial information. We first preprocess gene expression to make them amenable for wavelet filtering. We then use the Daubechies D4 Wavelet Transform \citep{daubechies1992ten} as wavelet filter.

\subsection{Gene Expression Over Location} \label{InputGeneration}
We leverage a pre-processing step, which we call \emph{input generation}, to combine spatial and expression data sources. We have both gene expression measurements and spatial coordinates for each sample. Each gene has an expression over each sample, and we can draw the sample over a 2D map by their coordinates. 

\begin{figure}[!htbp] 
\begin{center} 
\includegraphics[width=8cm]{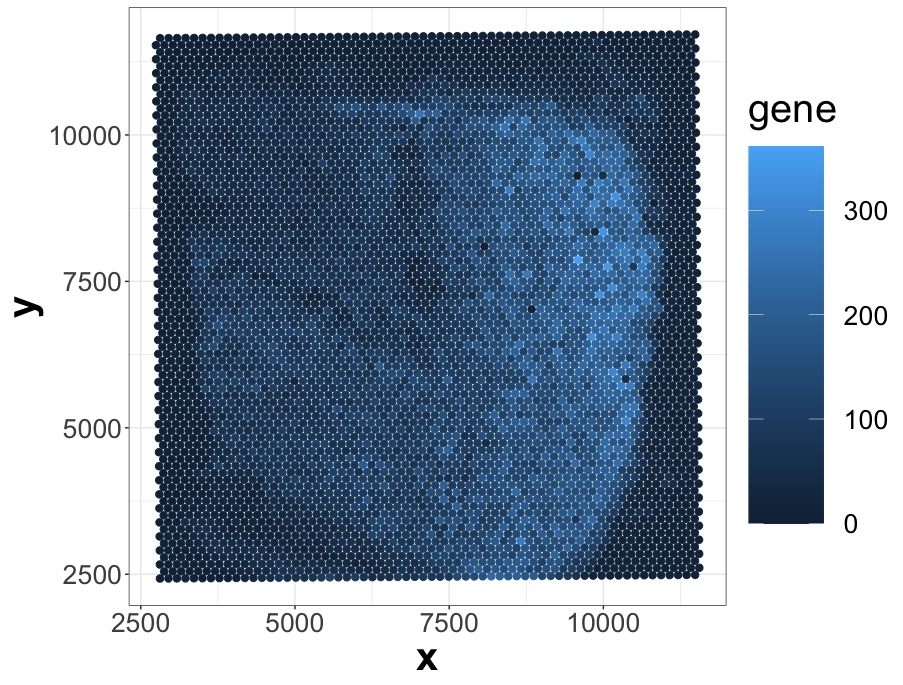}
\end{center}
\caption{The most dense gene has a structured expression pattern.} 
\label{STgeneExpression}
\end{figure}

As shown in Figure \ref{STgeneExpression}, the gene with the least sparsity in expression shows varying levels of expression across different spatial regions. Intuitively, we may consider the spatial expression pattern from this gene to be spatially related. However, most genes are sparse and do not show fluctuations in gene expression. Therefore, we first filter out those genes using a \texttt{kOverA} filter: we only keep genes that have an expression measure above $A$ in at least $k$ samples \citep{genefilter}. This has the effect of removing genes that are rarely active, though it is possible that strongly expressed genes still show no spatial relationships.

\begin{figure}[!htbp] 
\begin{center} 
\includegraphics[width=10cm]{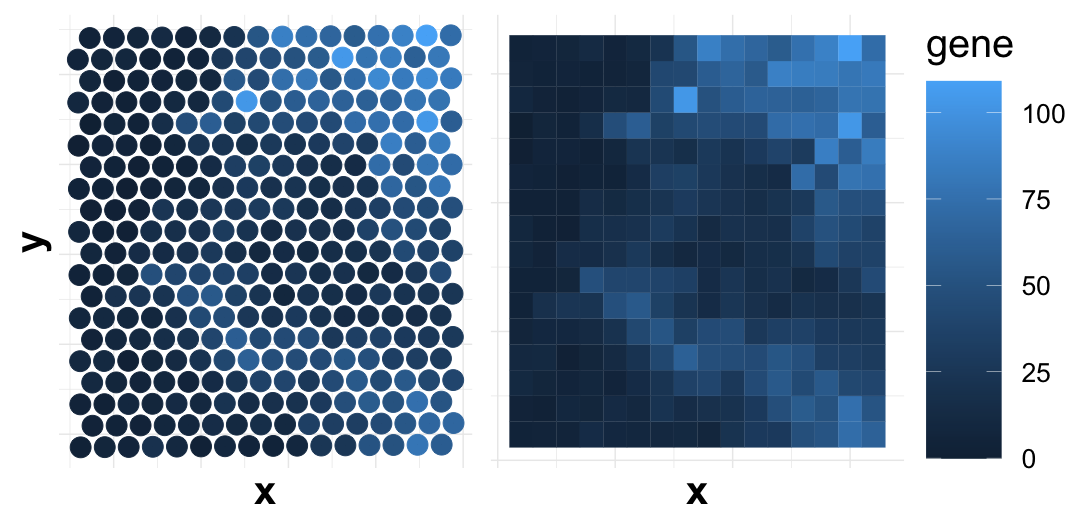}
\end{center}
\caption{Left: The gene expression over scattered samples over locations. Right: The local average of gene expression over grid locations, used as input to our algorithms.} 
\label{localAverage}
\end{figure}

We want gene expression data to have a neat grid structure that can be expressed as a matrix form. However, at first, expression measurements are roughly staggered over the spatial locations, as shown in Figure \ref{localAverage}. This is a consequence of the sampling strategy adopted by the 10x genomics Visium platform. To obtain an evenly resampled version of the expression pattern, we divide the two-dimensional space into several partitions and compute the local average of each partition. Detailed are shown in Algorithm \ref{pre}.
Each gene has an expression over the grid and we now can use a matrix $Z \in \mathbb{R}^{D^2 \times P}$ to represent it. The matrix $Z$ becomes the new input for analysis. We then obtain the same formulation as in Section \ref{problemSetup}, where we take $n = D^2$ and each location corresponds to a new observation. We have $\*f_{k} \in \mathbb{R}^{D^2}, k = 1,\ldots,K$ as the factor genes, computed by vectorizing the gene expression matrix.

The choice of $D$ depends on the wavelet scale used in the next section. We choose $D$ to be dyadic to prevent handling edge cases in the wavelet transformation (recalling Section \ref{waveBackground}). The dyadic size is also a natural choice in image analysis. In particular, we choose $D = 2^J$, where $J$ is the level of scale in the wavelet transformation. Larger $J$ allows finer recovery, but also requires more parameters and may result in overfitting.

\subsection{Wavelet Transformation and Shrinkage}
We apply a wavelet transformation with shrinkage to denoise the gene expression matrix and smooth observed spatial expression patterns. In simulations, we find that this technique gives more accurate recovery as the signal-to-noise ratio decreases and yields less noisy visualizations for the spatially-related genes.

We implement wavelet transformation and shrinkage on the processed data matrix $Z \in \mathbb{R}^{D^2 \times P}$. Each column of $Z$ is a vectorized expression matrix. For each column, we first reshape it into a $D\times D$ matrix and apply the wavelet transformation. We have a coefficient list with $W$ coefficients associated with each gene. Then we conduct wavelet shrinkage on the coefficient list using the threshold strategies described in Section \ref{waveBackground}. Finally, we vectorize the coefficient list of each gene into a length $W$ vector for each gene. Stacking all vectors together, we have a coefficient matrix $C \in \mathbb{R}^{W \times P}$. Note that the wavelet scale is specified by the size $D$ of the input matrix, i.e., we have $D = 2^J$, where $J$ is the scale. The details are specified in Algorithm \ref{waveletTrans}.

\subsection{Matrix Decomposition} \label{subsec_decomp}
The matrix of the shrunk wavelet coefficients $C$ gives a summary of the denoised matrix. This allows improved reconstruction of gene expression data. To obtain a low-dimensional approximation, we apply matrix factorization on the coefficient matrix $C$ after transformation. 
We use the same notation in Section \ref{problemSetup}, with the subscript $c$ to denote the decomposition on coefficient matrix $C$.

The resulting singular vectors can be used to estimate spatially structured factor genes (Section \ref{subsec_factorgene}). We use SVD as a frequentist approach to estimating $\hat{F_c}$ and $\hat{L}_c$. We also conduct EBMF (Section \ref{EBMF}), using the posterior expectation of $F_c$ and $L_c$ to estimate $\hat{F_c}$ and $\hat{L_c}$. EBMF can select the number of factors $K$ by itself, whereas $K$ must be manually specified in the SVD. The choice of $K$ is informed by inspecting the scree plot, as in spectral clustering and PCA. We choose the number of factors when current factors explain sufficient information and the diminishing returns of additional factors are no longer worth the additional cost. 
Given $\hat{F}_{c}$ and $\hat{L}_{C}$, we can compute the estimated coefficient matrix as $\hat{C} = \hat{F}_c\hat{L}_c^T$.

\subsection{Inverse Wavelet Transformation}

To transfer the coefficient matrix $\hat{C}$ back to the estimated location-by-gene matrix $\hat{Z}$, we apply the inverse wavelet transformation on columns of $\hat{C}$, each of which is a vectorized coefficient list for one gene. This results in a $D^2$ expression matrix for each gene. Then we vectorize the matrix and stack all vectors together. This yields the reconstructed matrix $\hat{Z}$. Details are given in Algorithm \ref{invWaveletTrans}.
For visualization, we also conduct a similar process for each column of $\hat{F_c}$. In this case, each column is associated with an factor gene. By applying the inverse wavelet transformation to each factor gene, we can build spatial gene expression matrices $M_1, \ldots, M_K$ representing gene factors.

\begin{algorithm}
\caption{Input Generation}\label{pre} 
\begin{algorithmic}[1] 
\Require Sample by gene matrix $Y \in \mathbb{R}^{N \times P}$, spatial matrix $S \in \mathbb{R}^{N \times 2}$, size of gene expression matrix $D$
\State Compute the range of $x,y$ coordinates from $S$, compute the coordinates of vertices of big rectangle map $B$ cover all $N$ samples spatially.
\State Partition interval $x$ and interval $y$ into $D$ equal length interval, together get $D^2$ partitions over rectangle $B$.
\While{$i \le P$}
\Comment{Consider gene $i$ expression over map $B$}
      \State Select $i$-th column of gene matrix $Y$
      \State Compute the local average of gene expression of gene $i$ in each partition
      \State Get $D^2$ matrix $G_i$ as gene expression of gene $i$
      \State Vectorize $G_i$ into a vector $g_i$ with length $D^2$
   \EndWhile
   \State Stacking all $\{g_i\}_{i=l}^P$ together into matrix $Z \in \mathbb{R}^{D^2 \times P}$
\State \textbf{return} matrix $Z$
\Comment{A transformed gene expression matrix}
\end{algorithmic}
\end{algorithm}

\begin{algorithm}[!htbp] 
\caption{Wavelet Transformation and Shrinkage}\label{waveletTrans} 
\begin{algorithmic}[1] 
\Require Location by gene matrix $Z \in \mathbb{R}^{D^2 \times P}$, threshold method, optional threshold parameter $\tau$. 
\Comment{Apply threshold on wavelet coefficient if threshold method is specified}

\While{$i \le P$}
\Comment{Consider gene $i$ expression as matrix $G_i$}
      \State Select $i^{\text{th}}$ $g_i$ column of gene matrix $Z$
      \State Form $g_i$ into expression matrix $G_i$ with size $D^2$
      \State Apply 2-D discrete wavelet transformation over $G_i$, get coefficient list $C_i$, with number of coefficients $W$
      \State Apply wavelet shrinkage over $C_i$ with optional parameter $\tau$
      \State Vectorize $C_i$ into a long vector $c_i$ with length $W$
   \EndWhile
   \State Stacking all $\{c_i\}_{i=l}^P$ together into matrix $C \in \mathbb{R}^{W \times P}$
\State \textbf{return} coefficient matrix $C$
\Comment{Column $i$ of $C$ store the coefficient of from gene $i$}
\end{algorithmic}
\end{algorithm}

\begin{algorithm}[!htbp] 
\caption{Inverse Wavelet Transformation}\label{invWaveletTrans} 
\begin{algorithmic}[1] 
\Require (reconstructed) Coefficient matrix $C$

\While{$i \le P$}
      \State Select $i^{\text{th}}$ $c_i$ column of gene matrix $C$
      \State Form $c_i$ into coefficient list $C_i$ with number of coefficients $W$
      \State Apply 2-D inverse wavelet transformation over $C_i$, get post-wavelet expression matrix $\hat{G_i}$
      \State Vectorize $\hat{G_i}$ into a long vector $\hat{g_i}$ with length $D^2$
   \EndWhile
   \State Stacking all $\{\hat{g_i}\}_{i=1}^P$ together into matrix $\hat{Z} \in \mathbb{R}^{D^2 \times P}$
\State \textbf{return} matrix $\hat{Z}$
\Comment{Post processing reconstructed gene expression matrix}
\end{algorithmic}
\end{algorithm}

\subsection{Evaluation} \label{evaluation}
We evaluate our proposal both quantitatively and qualitatively. Our quantitative results measures reconstruction error between the estimated $\hat{Z}$ and $Z$ using the Frobenius norm $\|\hat{Z} - Z\|_{F}^2$.

This evaluation is also used for hyperparameter tuning, such as the scale $J$ of the wavelet and the choice of wavelet thresholding method. Our qualitative result is given by visualization of the estimated factor gene expression matrices, with emphasis on capturing global spatial structure.

We use cross-validation in computing reconstruction error and calculating gene-wise errors. This evaluation is helpful in selecting spatially related genes. We also found a simple connection between the gradient of gene expression and spatial contribution, discussed in Section \ref{sectionRealData}. This discussion requires a measure of spatial expression smoothness. To this end, a simple step computing the fluctuation of gene expression would give a spatial gene selection that coincides with the reconstruction error selection. For calculating successive differences, consider $Z$ as the input matrix and let $\delta_{jk} = (z_{j,k+1} - z_{jk})^2, k = 1,\ldots,D^2$. Then we define the gradient by $\sum_{jk}\delta_{jk}$.

\section{Simulation} \label{Sec_sim}
We setup simulations to see whether representing spatial structure with a wavelet basis supports denoising and visualization of gene expression. We will see improved recovery in the low signal-to-noise ratio regime. Qualitatively, we also find factor gene visualizations to be more spatially consistent.

\begin{figure}[!htbp] 
\begin{center} 
\includegraphics[width=10cm]{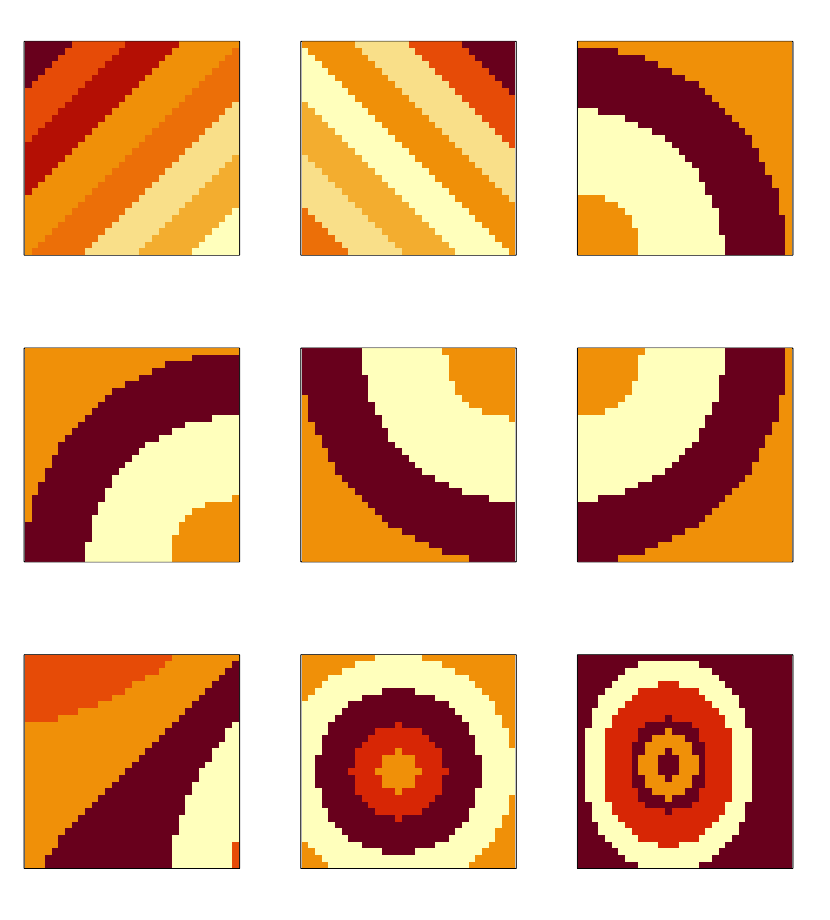}
\end{center}
\caption{The gene expression pattern for nine factor genes in the simulation experiment.} 
\label{truegenepattern}
\end{figure}

We first describe the simulation mechanism.
We the set number of factors to $K = 9$. We generate $K$ gene expression matrices $M_1, \ldots, M_K \in \mathbb{R}^{D^2}$, each representing a factor gene. These factor gene expression patterns are shown in Figure \ref{truegenepattern}.
We vectorize the patterns $M_{k}$ into factor genes $\*f_{k}$, which are then scaled so $\|\*f_{k}\| = 1$. 
We obtain $F \in \mathbb{R}^{D^2 \times K}$ by stacking all the $\*f_k$ together. We generate loadings $\*l_1, \ldots, \*l_K \in \mathbb{R}^{P \times K}$ by drawing coordinates independently from $N(0,1)$. We set $D = 32$ and wavelet scale $J = 5$.
We similarly stack $\*l_{k}$ to obtain $L \in \mathbb{R}^{P \times K}$. We use $Z = FL^T$ as the ground-truth signal matrix and add noise $E$ to yield data matrix $Z_d = FL^T + E$, where entries in $E$ are zero-mean normal noise that corrupt the underlying signal. 
In our analysis, we use a Daubechies D4 Wavelet as the wavelet filter, and for coefficient shrinkage we use hybrid thresholding.

We use the pipeline from Section \ref{methods}. The generated data $Z_d$ already reflects the structure produced by the processing of Section \ref{InputGeneration}. 
For comparison, we also directly decompose the matrix $Z_d$ with the SVD without any wavelet transformation. We call the resulting factors $\hat{F}_{raw}$ and reconstruction $\hat{Z}_{raw}$. We also have the same quantitative reconstruction error and qualitative visualization of factor gene $\hat{F}_{raw}$.
We also measure the size of the gradients across the estimated gene expression matrix $\hat{M}_i$. The gradient is computed by successive difference of neighboring image pixels. We calculated the sum of the squares of the gradient. This measurement shows whether the gene expression matrix has been smoothed. This property is of interest, since smoother estimates are often more visually appealing.

\begin{figure}[!htbp] 
\begin{center} 
\includegraphics[width=17 cm]{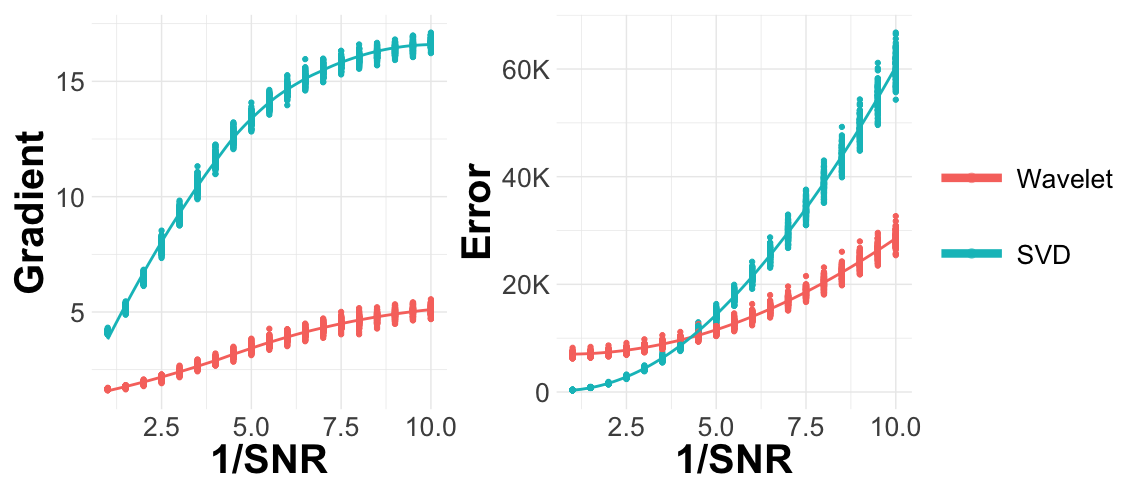}
\end{center}
\caption{The reconstruction error and gradient for estimated gene-by-location matrix with different SNR. The $x$-axis shows the rate of $\frac{1}{\text{SNR}}$ The left subplot shows the gradient changes. The gradient of gene expression is always lower with the wavelet technique, a consequence of its smoothing property. The right subplot shows the error -- the wavelet method has lower error as the magnitude of noise increase.} 
\label{error_gd_SNR}
\end{figure}

We denote the signal-to-noise ratio as $\text{SNR} = \frac{\text{sd}(Z)}{\text{sd}(E)}$, where $sd$ stands for standard deviation. Let $r = \frac{1}{\text{SNR}}$. We specify 19 evenly spaced settings of $r$ from 1 to 10. For each setting, we run 100 replicates with different simulated $Z_d$ and apply wavelet and SVD-based dimensionality reduction. 
The resulting average errors across $r$ are shown in the Figure \ref{error_gd_SNR}. The wavelet and shrinkage technique has better performance when $r$ is larger than 5, i.e., the low signal-to-noise ratio regime. The gradient of the gene expression matrix under two methods is shown in Figure \ref{error_gd_SNR}. The wavelet and shrinkage approach smooths edges in the factor gene expression image, giving a more interpretable visualization and lower error in this low SNR setting.

\begin{figure}[!htbp] 
     \centering
     \begin{subfigure}[b]{0.49\textwidth}
         \centering
         \includegraphics[width=\textwidth]{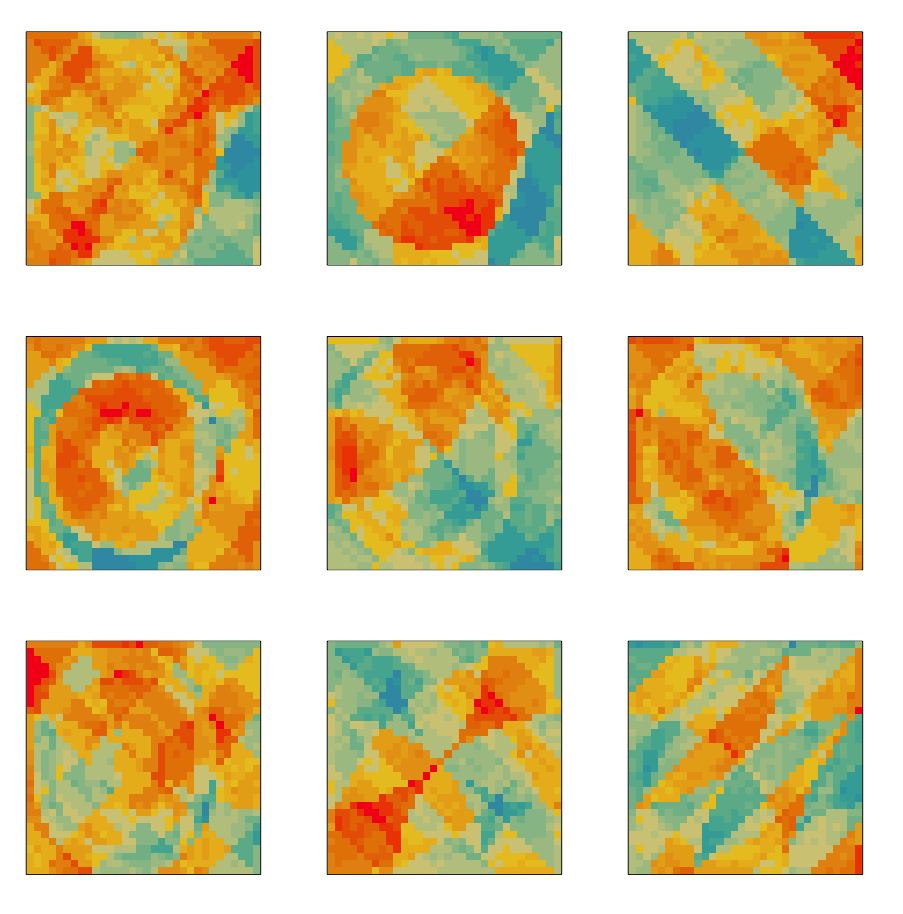}
         \caption{Factor genes without wavelet smooth}
     \end{subfigure}
     \hfill
     \begin{subfigure}[b]{0.49\textwidth}
         \centering
         \includegraphics[width=\textwidth]{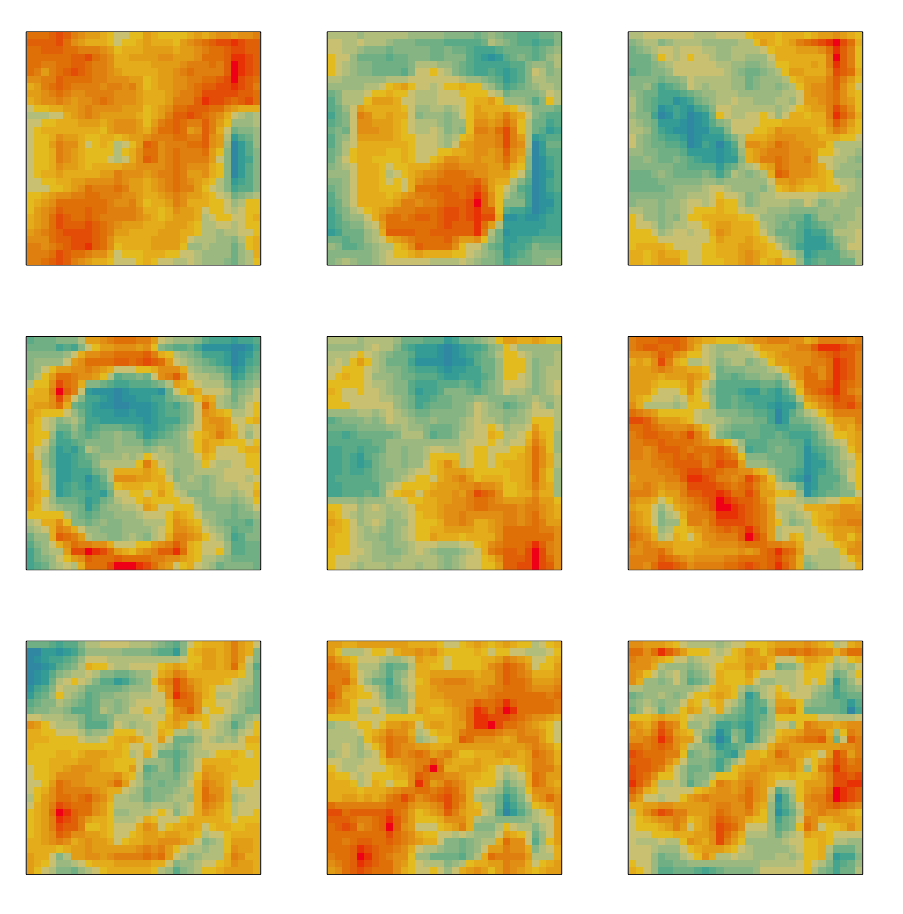}
         \caption{Factor genes with wavelet smooth}
     \end{subfigure}
        \caption{Factor gene visualization. Figure (a) implements SVD on $Z_d$, the visualization of column of $\hat{F}_{raw}$. Figure (b) implements SVD on coefficient matrix, the visualization of $\hat{F}$.}
        \label{eigengene}
\end{figure}

The factor genes are shown in Figure \ref{eigengene}. The SVD without wavelet technique has erratic, outlying pixels, especially in the last three genes. The visualization is sensitive to outliers. In contrast, the SVD combined with wavelet has smoother patterns. Like the SVD-based method, it appears to have mixed several of the true underlying factors in each of the recovered ones. Moreover, the sharp boundaries visible in the SVD factors become smoothed over in the wavelet-decomposition. The wavelet method applies a decomposition on coefficients space after thresholding, while SVD operates on individual pixels. The SVD capture more information, but also emphasize nuisance information induced by errors. Wavelet method also reduces model complexity, improving estimation accuracy. Other non-parametric methods, such as Fourier transformation and Gaussian Process-based methods also operate on coefficients space. Still, they would struggle to capture sharp transitions, since their bases are smooth functions.

Nonetheless, both the SVD and wavelet-based visualizations reflect spatial trends in the true factor genes (in Figure \ref{truegenepattern}). 

\newpage

\section{Real Data Experiment} \label{sectionRealData}
In this section, we show that wavelet and shrinkage technique reduces reconstruction error quantitatively. We ran our method on a public spatially resolved transcriptomics data \citep{STdata}. The dataset represents a single biological sample from the human brain dorsolateral prefrontal cortex (DLPFC) region, measured with the 10x Genomics Visium platform.
Further, we identify a simple connection between the gradient of gene expression and quantitative error. A simple step computing the fluctuation of gene expression alone (calculating successive difference of the gene expression image) selects genes that have reduced reconstruction error when using wavelet-guided dimensionality reduction.

\begin{figure}[!htbp] 
\begin{center} 
\includegraphics[width=10cm]{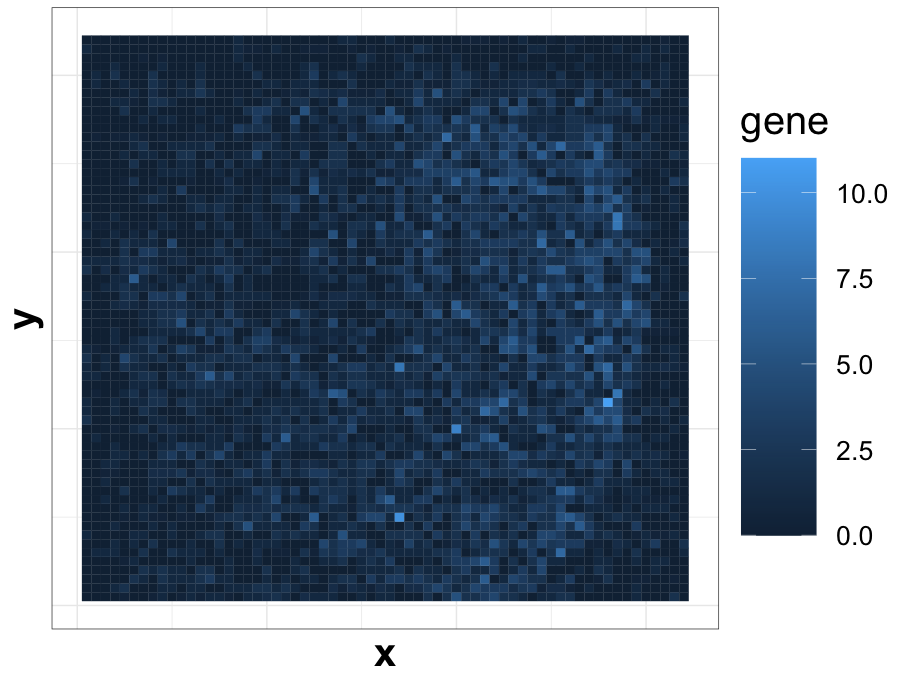}
\end{center}
\caption{Gene expression over the grid.} 
\label{STAverage}
\end{figure}

We first process the ST data through our pipeline. The ST data contains 4992 observations with 33538 genes. Expressions from most genes are sparse. We implement the pipeline from Section \ref{methods}. We pre-process as in Section \ref{InputGeneration} and then apply a \texttt{kOverA} filter to select genes. We find $k = 3,A = 7$ gives us 721 genes with average expression as 2.71. Then we apply Algorithm \ref{pre} to transfer the sample-by-gene matrix to a grid-by-gene matrix. We set $D = 64$ and wavelet scale $J = 6$. 
We obtain an image heatmap for each gene like in Figure \ref{STAverage}.

We then vectorize $P = 721$ image expression and stacking vectors together, we have generated input data $Z \in \mathbb{R}^{64^{2} \times 721}$.
We run our pipelines with and without wavelet transformation for evaluation. We first implement SVD and EBMF on $Z$ or on coefficient matrix $C$ after applying wavelet transformation and thresholding each column. We choose the number of factors $K$ by examining singular values. If the singular value is larger than 500, we keep it as a factor. In EBMF, we set upper bound that $K$ to be smaller than the corresponding $K$ in SVD, then let algorithm choose $K$ itself.

Quantitative evaluation metric is conducted by cross-validation on non-zero entries. We set 5 folds cross-validation with replacement. In particular, we select random $\frac{1}{5}$ non-zero entries in matrix $Z$ and set them to zero, then we save matrix as the masked matrix (train data) $Z_{train}$. We store the values and position of the masked entries as test data $Z_{test}$. We then ran methods on $Z_{train}$. The result shows a difference in whether to use the wavelet technique. The result from SVD and EBMF are close to each other coordinate-wise, hence we only show one of them in some results below.We include the comparison between SVD and EBMF in the Appendix \ref{App_comparison}.
As in Section \ref{Sec_sim}, we have $\hat{F}$ and $\hat{Z}$ from the wavelet approach and $\hat{F}_{raw}$ and $\hat{Z}_{raw}$ on the original data. 

\subsection{Total Error and Parameter Tuning}

\begin{figure}[!htbp] 
\begin{center} 
\includegraphics[width=13cm]{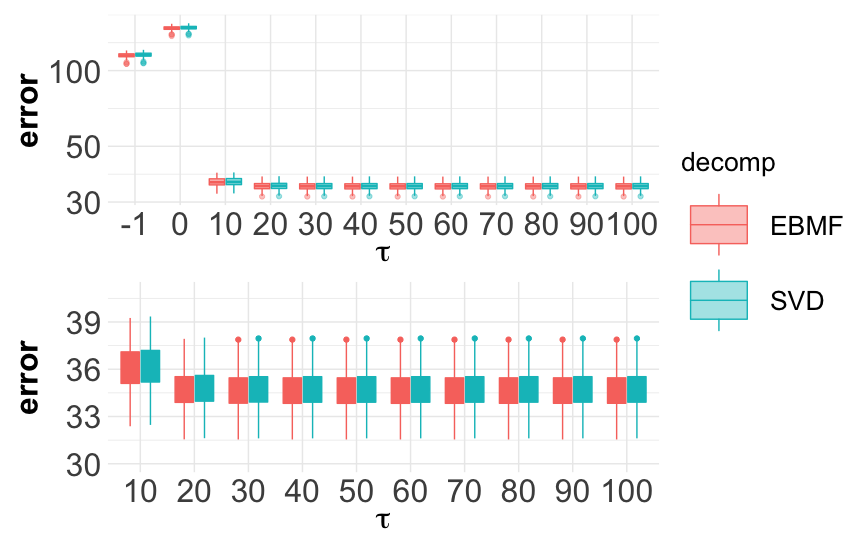}
\end{center}
\caption{The reconstruction error for different wavelet parameters. The upper plot contains the result from all settings, the leftmost $\tau = -1$ result is the decomposition without wavelet transformation. $\tau = 0$ indicates the hybrid thresholding. The bottom plot zooms into the experiment under manual threshold for $\tau \in \left[10, 100\right]$.} 
\label{totalError}
\end{figure}

We compute the reconstruction loss,
$$
\frac{1}{N}\sum_{i,j \in \text{test}}(\hat{Z}_{ij} - Z_{ij})^2
$$
where $N$ is the number of test entries. We evaluate the loss only on test entries. We first tune parameters. We set up three settings: decompose the raw data with SVD or EBMF; wavelet transformation with hybrid thresholding; wavelet transformation with manual thresholding with threshold $\tau = 10,20,\ldots, 100$. In each setting, we ran 100 replicates. The reconstruction error shown in Figure \ref{totalError}. 

As shown in the upper panel of Figure \ref{totalError}, wavelet thresholding with manually set $\tau$ reduces error compared to decomposing raw data. The wavelet has the most positive effect when $\tau = 40$, as shown in the bottom panel. We use $\tau = 40$ in our following analysis.

\subsection{Genewise Error}

We then evaluate how method performance varies for each gene by calculating the genewise reconstruction error. This reveals genes whose strong spatial expression structure leads to improved performance when using a wavelet basis. We still hold out $\frac{1}{5}$ of the entries at random as a test set. We compute the reconstruction error of test entries on each column of $\hat{Z} - Z$. We calculate entry-wise loss across 100 replicates and estimate the average loss.
We compare genewise errors with or without wavelet transformation. The SVD and EBMF show the same result regarding the decomposition method. We show the result of EBMF in Appendix \ref{App_comparison} Figure \ref{EBMF_error_per_gene}.

\begin{figure}[!htbp] 
\begin{center} 
\includegraphics[width=10cm]{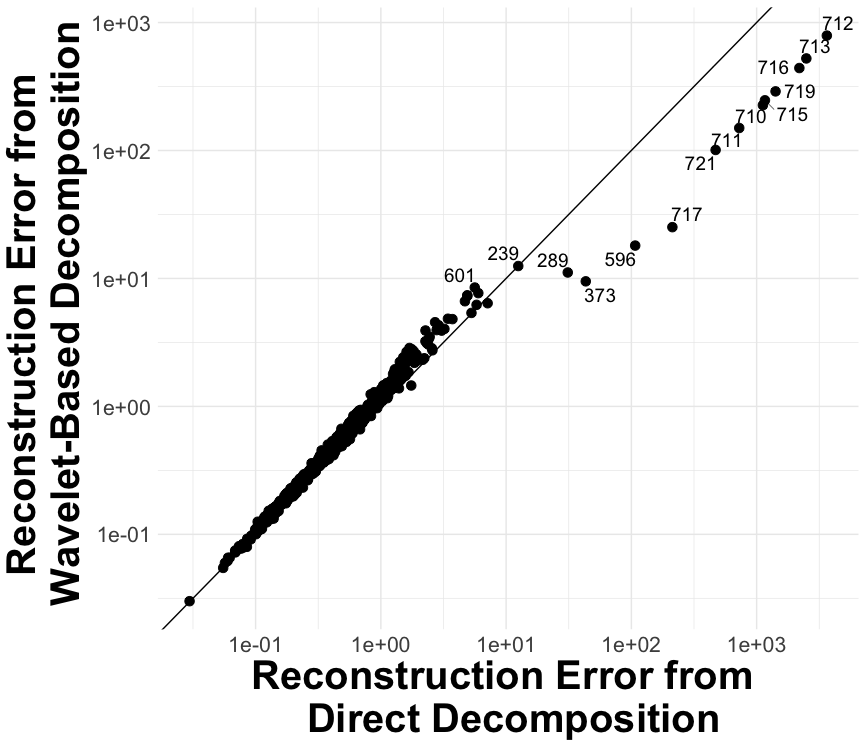}
\end{center}
\caption{The reconstruction error for each gene. The $x$-axis is the error from EBMF on raw data, the $y$-axis is the error from the combination of wavelet thresholding and EBMF.}
\label{EBMF_error_per_gene}
\end{figure}

\begin{figure}[!htbp] 
\begin{center} 
\includegraphics[width=13cm]{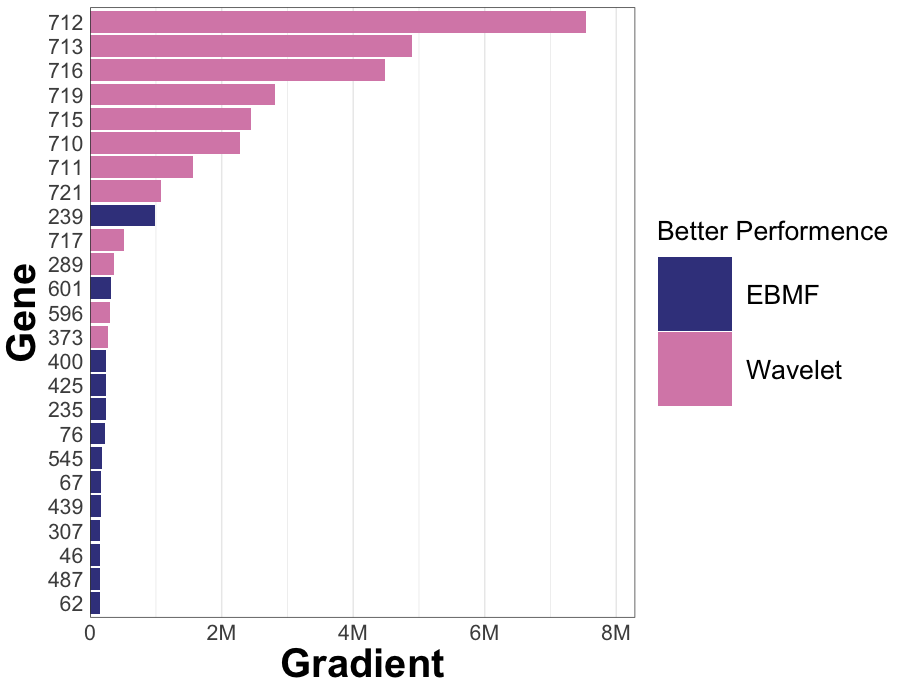}
\end{center}
\caption{The sum of squares of the gradient for each gene. The $y$-axis is the index of each gene. Each bar's color shows whether that gene has better performance when using a wavelet basis. The M on the $x$-axis stands for "millions".}
\label{gd_per_gene}
\end{figure}

As we can see, most of the genes have lower reconstruction error when directly applying the SVD or EBMF.  However, for some genes, wavelet smoothing reduces error. For example, this is seen in genes 712, 713, 716, 719, 715, 710, 711, 721, 717, 596, 373, and 289. We conjecture that these genes have acute fluctuations as well as clear spatial patterns. The expression matrix of these genes would have a larger gradient and distinct edges. To verify our conjecture, we calculated the sum of squares of the gradient of each gene expression matrix. We show the result in Figure \ref{gd_per_gene}.

The result verifies our conjecture: genes with larger gradients have better reconstruction under the wavelet-guided decomposition. This suggests a pre-processing step for selecting wavelet-suited genes by calculating the gradient of each gene. 
The spatially related genes would have a lower reconstruction error. Among these spatially related genes, 
one possibility is that we can divide them into two groups, one for decomposition directly and the other for the wavelet technique.

\subsection{Factor Genes}
\begin{figure}[!htbp] 
     \centering
     \begin{subfigure}[b]{0.49\textwidth}
         \centering
         \includegraphics[width=\textwidth]{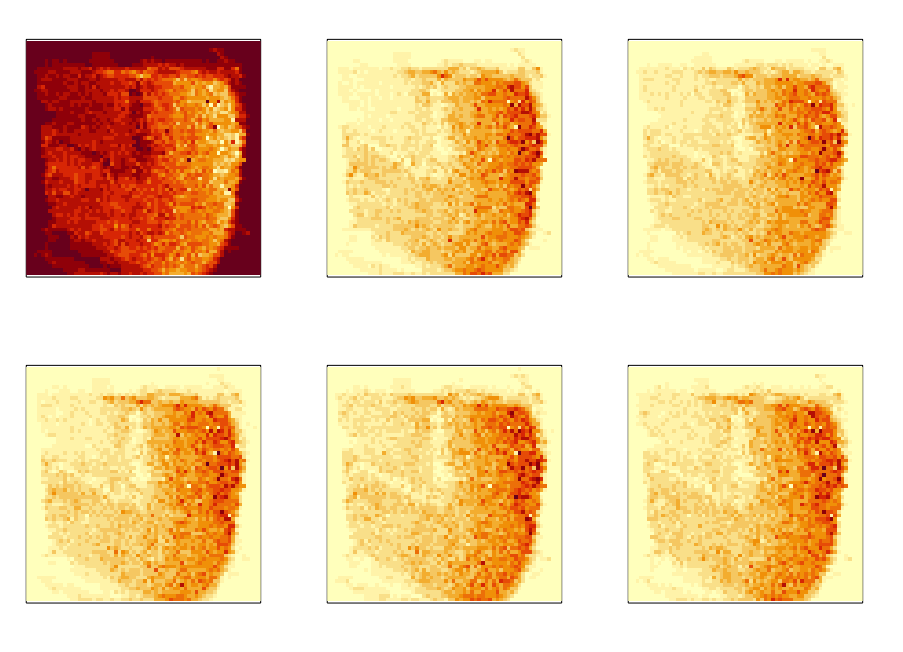}
         \caption{Factor genes without wavelet smooth}
     \end{subfigure}
     \hfill
     \begin{subfigure}[b]{0.49\textwidth}
         \centering
         \includegraphics[width=\textwidth]{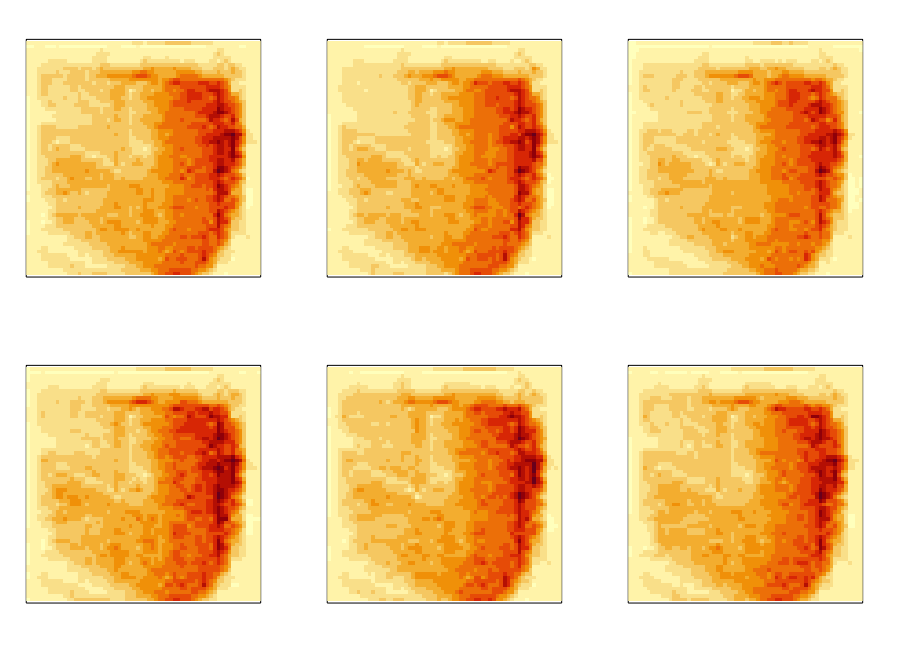}
         \caption{Factor genes with wavelet smooth}
     \end{subfigure}
        \caption{The top left figure is first factor gene, the following figure by row is genes with largest loadings on first factor gene: gene 712, 713, 716, 719, 715.}
        \label{raw_wave_eigengene1}
\end{figure}

Now we visualize the top factor genes and genes with high loadings on these factor genes. In Figure \ref{raw_wave_eigengene1}(a), we decompose $Z$ using the SVD and plot the first factor gene and matrix slides of genes with the largest 5 loadings on that factor. The factor gene captured the same patterns as the original genes. To improve visualization, we find the analogous wavelet-based factors (we use manual thresholding with $\tau = 40$), shown in Figure \ref{raw_wave_eigengene1}(b).

\begin{figure}[!htbp] 
     \centering
     \begin{subfigure}[b]{0.49\textwidth}
         \centering
         \includegraphics[width=\textwidth]{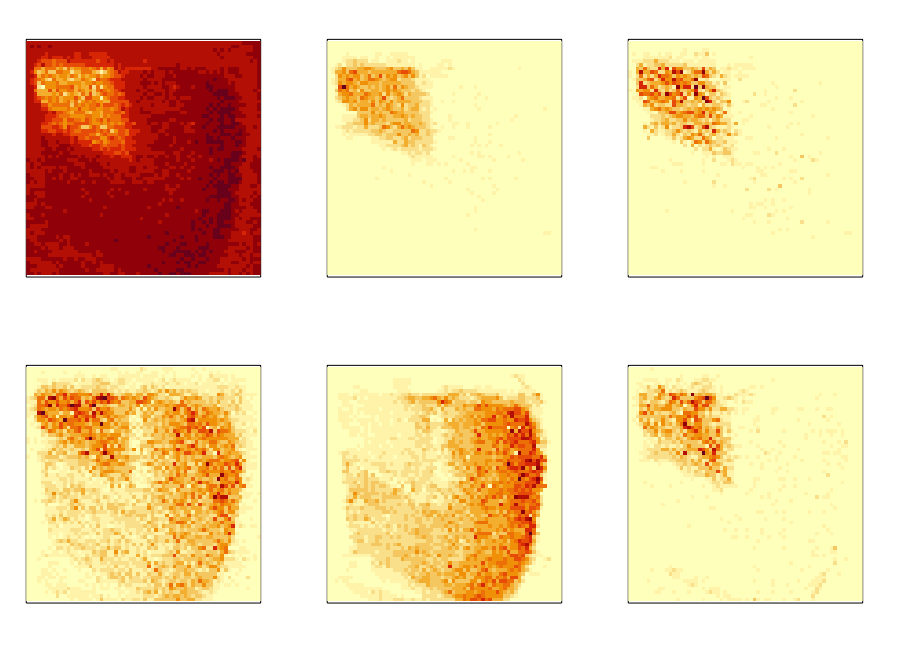}
         \caption{Factor genes without wavelet smooth}
     \end{subfigure}
     \hfill
     \begin{subfigure}[b]{0.49\textwidth}
         \centering
         \includegraphics[width=\textwidth]{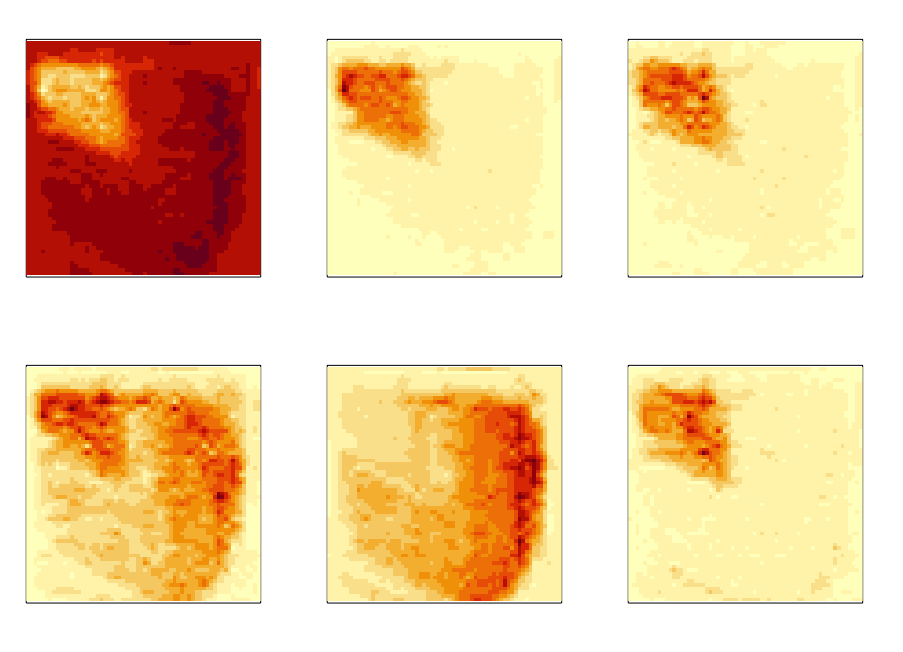}
         \caption{Factor genes with wavelet smooth}
     \end{subfigure}
        \caption{The top left figure is the second factor gene, the following figure by row is genes with largest loadings on second
        factor gene: gene 596, 289, 373, 712, 578.}
        \label{raw_wave_eigengene2}
\end{figure}

The wavelet thresholding approach smooths over edges in the original visualization. The first factor gene captures the global spatial expression. We have a similar result for the second factor gene, as shown in Figure \ref{raw_wave_eigengene2}.

The second factor gene is orthogonal to the first one, capturing different spatial structures in gene expression. Different factor genes capture global and local structure, and using a wavelet decomposition provides denoised spatial expression visualization.

In conclusion, we can use this dimensionality reduction technique for spatial gene selection and extraction. We select wavelet-suited genes based on the calculated gradient. Then, we can select spatially related genes based on reconstruction error via cross-validation. Alternatively, we can extract factor gene to capture spatial information and use factor genes for visualization and further analysis.

\subsection{Code Vignette}
We provide a small code block to show vignette of generating Figure \ref{raw_wave_eigengene1} and \ref{raw_wave_eigengene2}, shown in code block \ref{code}.
\begin{lstlisting}[language=R, caption=The basic wavelet-based dimensionality reduction workflow. \texttt{kOverA\_ST} transforms the original spatial expression context into an image. This is processed by \texttt{waveST}. The final dimensionality reduction step is performed by \texttt{decompose}., label = code] 
library(waveST)
res = kOverA_ST(k = 3,A = 7)
waveST = waveST(data = res$df, spatial = res$viz)

#### Figure 11
waveST = decompose(waveST, wavemethod = "raw", decom_method = "SVD", K = 5)
plot(waveST, k = 1, wave = FALSE)

#### Figure 12
waveST = decompose(waveST, wavemethod = "wave", decom_method = "SVD", K = 5, tau = 40)
plot(waveST, k = 1, wave = TRUE)
\end{lstlisting}

We have made a new package \texttt{waveST} containing the workflow we developed. The package is available at \url{https://github.com/OliverXUZY/waveST}. The \texttt{kOverA\_ST} function reduces data from an original \citep{STdata} class input using the \texttt{kOverA} technique. Then we use \texttt{waveST} function to construct a S4 class \texttt{waveST}, containing input generated by Algorithm \ref{pre}. This object-oriented approach stores properties of the original spatial experiment and simplifies downstream calls, like decomposition and visualization. We use the \texttt{decompose} function to apply all decomposition methods. In line 6, we apply the SVD to our original data, setting the number of factors to 5. In line 11, we apply the wavelet-based reduction and apply a manual threshold with $\tau = 40$. We use \texttt{plot} and \texttt{k=1} to visualize the first factor gene and matrix slides of the genes with the largest 5 loadings on that factor.

\section{Conclusion}
We have proposed a pipeline for dimensionality reduction that respects spatial structure. Both simulations and real data experiments demonstrate that wavelet and shrinkage techniques show positive results in spatially resolved transcriptomics data. We highlight the idea of combining image processing techniques and statistical methods for application in a spatial genomics context. One future direction is splitting genes into a groups suited or not for wavelet-based decomposition, and implementing decomposition with or without wavelet. Another direction is to focus input generation on only those genes that are thought to be spatially related.  For genes not related to spatial information, we may perform regular decomposition on original data $Y \in \mathbb{R}^{N \times P}$, abandoning the spatial information of those genes. We expect this to improve reconstruction performance. In further analysis, it is worth considering other wavelet smoothing techniques and wavelet filters. The current methods incorporate little biological information. Bringing more domain knowledge will require further techniques, but is expected to yield better results. The input generation computes local average over even grids; however, it is possible to apply the wavelet method for irregularly spaced data \citep{nason2008wavelet}. We hope wavelet methods will be useful in adapting existing methods for statistical genomics to the spatial setting.

\section*{Acknowledgments}
We thank Joseph Arthur for valuable discussions. 
This research was performed using the compute resources and assistance of the UW-Madison Center For High Throughput Computing (CHTC) in the Department of Computer Sciences. The CHTC is supported by UW-Madison, the Advanced Computing Initiative, the Wisconsin Alumni Research Foundation, the Wisconsin Institutes for Discovery, and the National Science Foundation, and is an active member of the OSG Consortium, which is supported by the National Science Foundation and the U.S. Department of Energy's Office of Science.

\bibliography{references}  

\newpage
\appendix
\section{Comparison Between SVD and EBMF} \label{App_comparison}
This section we showed the comparison between the reconstruction result between SVD and EBMF.

\subsection{Simulation}
In this section, we show the element-wise comparison between the reconstruction of SVD and of EBMF with wavelet technique.

\begin{figure}[!htbp] 
\begin{center} 
\includegraphics[width=10cm]{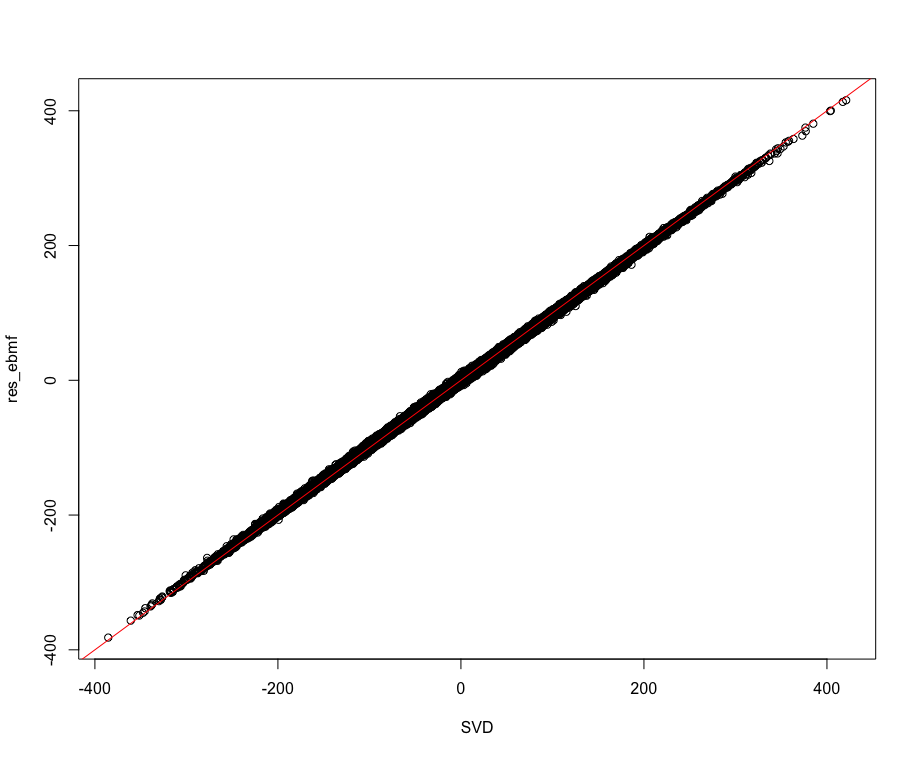}
\end{center}
\caption{The element-wise comparison of two reconstruction matrices. The $x$-axis shows the result of SVD decomposition, the $y$-axis shows the result of EBMF decomposition.}
\label{EBMF_error_per_gene}
\end{figure}

\subsection{Real Data Result}
In this section, we show the reconstuction error per gene result for SVD, similar to Figure \ref{EBMF_error_per_gene}.

\begin{figure}[!htbp] 
\begin{center} 
\includegraphics[width=10cm]{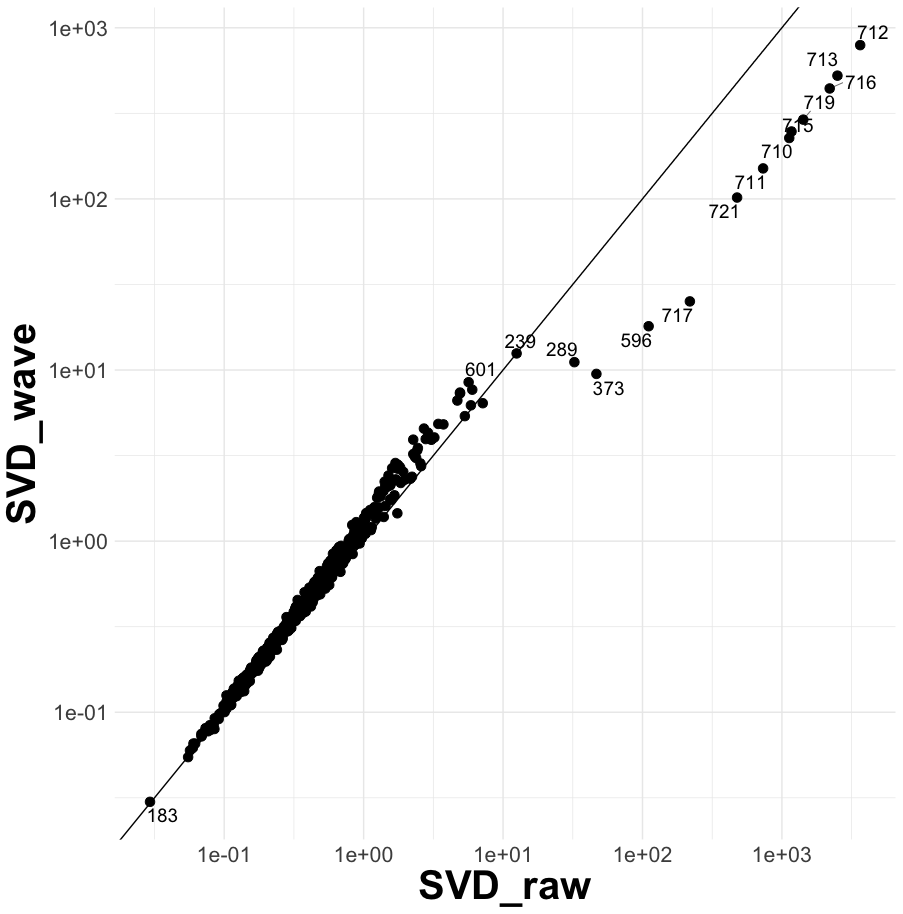}
\end{center}
\caption{The reconstruction error for each gene. The $x$-axis is the error from SVD on raw data, the $y$-axis is the error from the combination of wavelet thresholding and SVD.}
\label{EBMF_error_per_gene}
\end{figure}

\begin{figure}[!htbp] 
     \centering
     \begin{subfigure}[b]{0.49\textwidth}
         \centering
         \includegraphics[width=\textwidth]{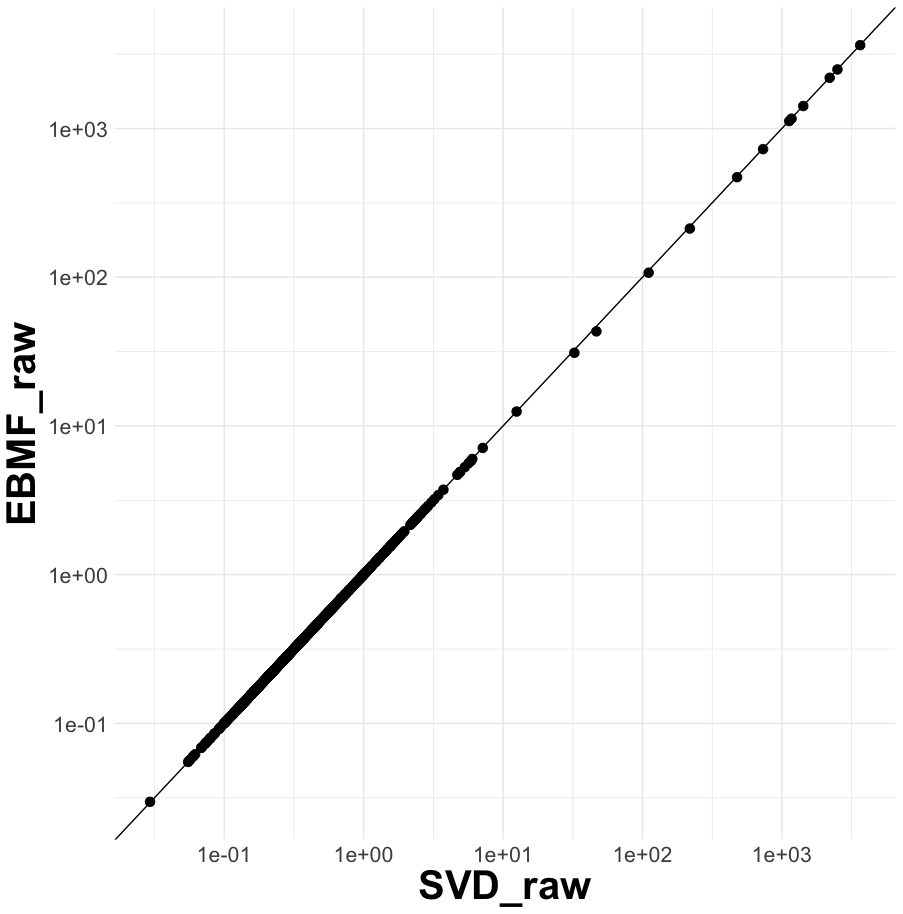}
         \caption{Error per gene without wavelet transformation.}
     \end{subfigure}
     \hfill
     \begin{subfigure}[b]{0.49\textwidth}
         \centering
         \includegraphics[width=\textwidth]{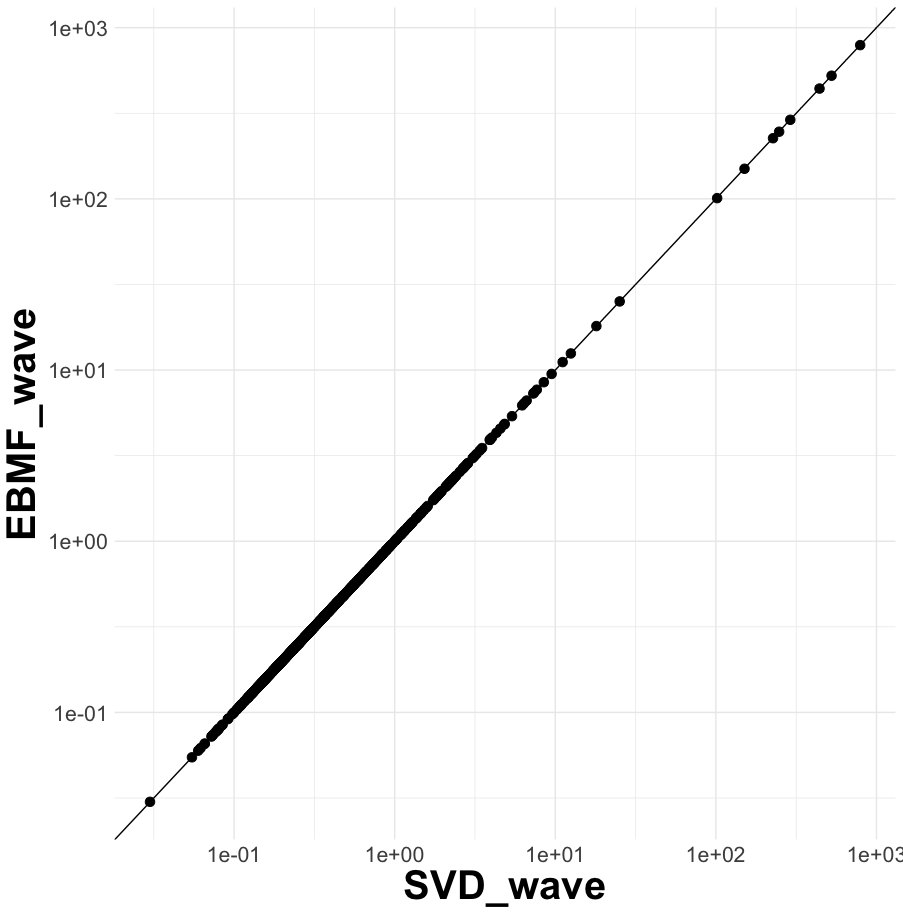}
         \caption{Error per gene with wavelet transformation.}
     \end{subfigure}
        \caption{The reconstruction error for each gene.}
        \label{raw_wave_error_per_gene}
\end{figure}

Then we show the result between SVD and EBMF with and without wavelet technique in Figure \ref{raw_wave_error_per_gene}. 

As we can see, SVD and EBMD have the very similar result.
\end{document}